\documentclass[journal=nalefd,manuscript=article]{achemso}

\usepackage{chemformula} % Formula subscripts using \ch{}
\usepackage[T1]{fontenc} % Use modern font encodings

\author{Loredana Ricciardi}
\affiliation{CNR NANOTEC- Institute of Nanotechnology U.O.S. Cosenza, 87036 Rende, (CS) - Italy}
\email{loredana.ricciardi@cnr.it}
\author{Sharmistha Chatterjee}
\affiliation{Department of Physics, University of Calabria, 87036 Rende, (CS) - Italy}
\author{Giovanna Palermo}
\affiliation{Department of Physics, University of Calabria, 87036 Rende, (CS) - Italy}
\alsoaffiliation{CNR NANOTEC- Institute of Nanotechnology U.O.S. Cosenza, 87036 Rende, (CS) - Italy}
\author{Elisabeta I. Szerb}
\affiliation{"Coriolan Dragulescu" Institute of Chemistry, Romanian Academy, 24 Mihai Viteazu Bvd., 300223 Timisoara - Romania}
\author{Alessia Sanna}
\affiliation{CNR NANOTEC- Institute of Nanotechnology U.O.S. Roma, 00185 Piazzale Aldo Moro 5 - Italy}
\author{Francesca Palermo}
\affiliation{Department of Physics, University of Calabria, 87036 Rende, (CS) - Italy}
\alsoaffiliation{CNR NANOTEC- Institute of Nanotechnology U.O.S. Roma, 00185 Piazzale Aldo Moro 5 - Italy}
\author{Nicola Pieroni}
\affiliation{CNR NANOTEC- Institute of Nanotechnology U.O.S. Roma, 00185 Piazzale Aldo Moro 5 - Italy}
\author{Michela Fratini}
\affiliation{CNR NANOTEC- Institute of Nanotechnology U.O.S. Roma, 00185 Piazzale Aldo Moro 5 - Italy}
\author{Roberto Bartolino}
\affiliation{CNR NANOTEC- Institute of Nanotechnology U.O.S. Cosenza, 87036 Rende, (CS) - Italy}
\author{Alessia Cedola}
\affiliation{CNR NANOTEC- Institute of Nanotechnology U.O.S. Roma, 00185 Piazzale Aldo Moro 5 - Italy}
\author{Massimo La Deda}
\affiliation{CNR NANOTEC- Institute of Nanotechnology U.O.S. Cosenza, 87036 Rende, (CS) - Italy}
\alsoaffiliation{Department of Chemistry and Chemical Technologies, University of Calabria, 87036 Rende, (CS) - Italy}
\alsoaffiliation{National Interuniversity Consortium of Materials Science and Technology, LASCAMM, Calabria Unit, 87039-Rende, (CS) - Italy}
\email{massimo.ladeda@unical.it}
\author{Giuseppe Strangi}
\affiliation{CNR NANOTEC- Institute of Nanotechnology U.O.S. Cosenza, 87036 Rende, (CS) - Italy}
\alsoaffiliation{Department of Physics, University of Calabria, 87036 Rende, (CS) - Italy}
\alsoaffiliation{Department of Physics and Case Comprehensive Cancer Center, Case Western Reserve University, 10600 Euclid Avenue, Cleveland, Ohio 44106 - USA}
\email{giuseppe.strangi@case.edu}

\title[An \textsf{achemso} demo]
  {Glioblastoma Treatments with Photo-Nanotherapeutics Induce Massive Devascularization and Tumor Elimination}

\abbreviations{IR,NMR,UV}

\begin{document}

%\begin{tocentry}
%\centering
%\includegraphics[width=9cm,height=6cm,keepaspectratio]{TOC}
%%
%\end{tocentry}

\begin{abstract}
  Glioblastoma multiforme (GBM) is one of the deadliest and most aggressive cancers, remarkably resilient to current therapeutic treatments. Here, we report \textit{in vivo} studies of GBM treatments based on photo-nanotherapeutics able to induce synergistic killing mechanisms. Core-shell nanoparticles have been weaponized by combining the photophysical properties of an Ir(III) complex - a new generation PDT agent - with the thermo-plasmonic effects of resonant gold nanospheres. To investigate the damages induced in GBM treated with these nanosystems and exposed to optical radiation, we recurred to the X-ray phase contrast tomography (XPCT). This high-resolution 3D imaging technique highlighted a vast devascularization process by micro-vessels disruption, which is responsible of a tumor elimination without relapse.
\end{abstract}

\section{Introduction}
Glioblastoma multiforme (GBM) is the most aggressive and prevalent brain cancer, accounting for over 70\% of high-grade gliomas diagnosed.\cite{louis20072007, schwartzbaum2006epidemiology, thakkar2014epidemiologic, hardee2012mechanisms} Characterized by necrotic primary tumor centers with an abundant and aberrant neovascularization,\cite{jain2007angiogenesis} current therapeutic treatments include surgical resection followed by adjuvant chemotherapy and radiotherapy.\cite{stupp2009effects}  Despite the recent advances in conventional therapeutic strategies, the GBM prognosis remains poor, with a median survival of 12-14 months.\cite{koshy2012improved,tran2010survival, legler1999brain,de2017nanomedicine} One of the important features of GBM cells is their infiltrative nature which leads to indistinguishable margins between normal and malignant brain tissue.\cite{silbergeld1997isolation} Consequently, a complete resection is rarely feasible. In order to identify and eliminate residual tumor cells at the boundaries of the resection area and at the same time minimize the damage to the surrounding healthy brain tissue, currently a new therapeutic strategy - undergoing phase I clinical trial - consists of intraoperative fluorescent-guided resection followed by photodynamic treatment.\cite{dupont2019intraoperative} Photodynamic Therapy (PDT) is an emerging cancer treatment approach, that employs light to trigger a photodynamic mechanism mediated by a drug - the photosensitizer (PS) - able to absorb and transfer the radiant energy to the ubiquitous molecular oxygen, generating reactive oxygen species (ROS) i.e. free radicals (type I reaction) and singlet oxygen - $^1$O$_2$ - (type II reaction).\cite{dolmans2003photodynamic,castano2004mechanisms} Recently, extraordinary efforts have been focused to investigate transition metal complexes (TMs) as PSs for PDT applications.\cite{mckenzie2019transition}  Because of their photo- physical properties, TMs meet several essential requirements as PDT agents.\cite{zhao2013triplet}  Furthermore, the remarkable luminescence associated with long emission lifetimes make TMs useful probe for imaging.\cite{baggaley2012lighting,ko2019dual} Currently, the application of these complexes in PDT is in its infancy, with the first TM entered into human clinical trials in early 2017.\cite{monro2018transition}

The development of nanotechnologies in the last decades has opened new promising avenues in the biomedical field, improving the clinical outcomes for many lethal diseases, such as cancer.\cite{norouzi2019clinical,wong2020nanomaterials} Among the wide variety of nanomaterials reported in the literature, inorganic nanoparticles possess very interesting properties for clinical application.\cite{giner2016inorganic,bobo2016nanoparticle} In this frame, gold-core silica shell nanoparticles can be used to vehiculate treatment agents in specific sites by enhancing significantly their efficacy, thereby proving to be extremely promising for theranostic purposes.\cite{moreira2018gold} Gold nanoparticles exhibit interesting size-dependent optical properties associated with the localized surface plasmon resonance phenomenon (LSPR).\cite{catherine2017gold, hutter2004exploitation}  The interaction with light at a specific resonance wavelength (which depends on the particle size and shape) leads to the absorption of the radiant energy and its conversion both into heat and scattered radiation. Accordingly, if properly localized inside or in close proximity of tumor cells, light-activated metal nanoparticles induce temperature increase with consequent photothermal-mediated cell ablation (Photothermal Therapy, PTT).\cite{huang2008plasmonic,jaque2014nanoparticles,hu2020reprogramming} The inclusion of a silica shell confers stability to the gold nucleus and provides a reservoir for the encapsulation of PSs molecules acting as PDT therapeutic agent and imaging probe, simultaneously, leading to the implementation of multifunctional nanoplatforms.\cite{vivero2012silica} Recently, we have reported the design and the development of a de novo nanosystem for simultaneous cellular imaging, photodynamic and photothermal therapies.\cite{ricciardi2017plasmon} The nanostructure has been realized by embedding a highly luminescent water soluble Ir(III) complex (\textbf{Ir$_1$})\cite{ricciardi2014ionic} into gold core-silica shell nanoparticles \textbf{Ir$_1$-AuSiO$_2$}. In vitro photo-cytotoxicity tests on GBM cells clearly demonstrated the potential of this nanoplatform to play a key role in the imaging and treatment of GBM. These results emphasized striking synergistic photodynamic and photothermal effects as a result of the coupling between the photophysical properties of the TM with the thermoplasmonic effects of the gold nanospheres. \cite{ricciardi2017plasmon}

Here, we report \textit{in vivo} results based onto the phototherapeutic treatment of human GBM xenograft mouse model after \textbf{Ir$_1$-AuSiO$_2$}  intratumoral injection, observing regression and complete tumor elimination without relapse. High-resolution X-ray phase contrast tomography (XPCT),\cite{zhang2020high,momose1996phase} was employed to image the 3D tumor vascular network with microscale resolution of  \textit{ex vivo} samples, shedding light on the mechanism underlying the tumor eradication. A schematic illustration summarizing the  \textit{in vivo} therapeutic approach is displayed in Figure 1.

\begin{figure}[H]
\includegraphics[width=0.9\columnwidth]{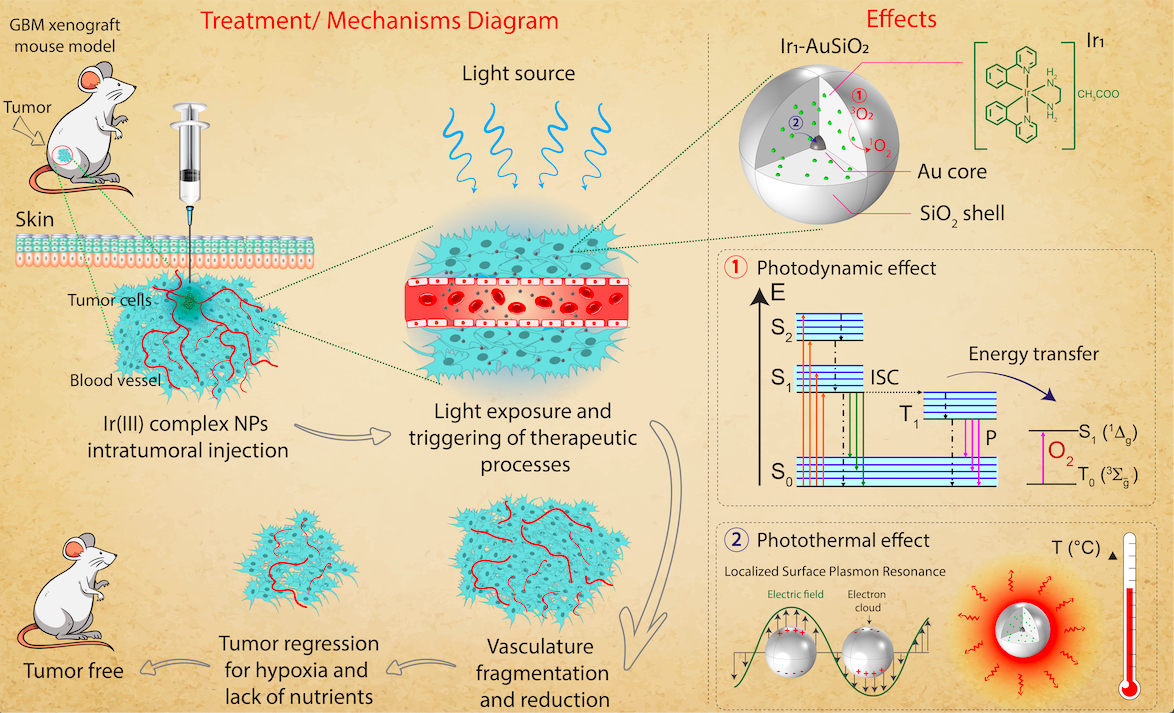}
\centering
\caption{\textbf{Overview of the therapeutic approach for human GBM treatment in xenograft mouse model.} Intratumoral injection of \textbf{Ir$_1$-AuSiO$_2$}, distribution and accumulation next to blood vessel walls followed by light-activation of photodynamic and photothermal processes. \textcircled{1} The light exposure excites the photosensitizer (\textbf{Ir$_1$}) loaded in the silica shell of \textbf{Ir$_1$-AuSiO$_2$} nanoparticles, from the ground state to the short-lived excited singlet state. Following intersystem crossing (ISC), the resulting triplet state can directly interact with molecular oxygen ($^3$O$_2$) and via an energy transfer mechanism generate cytotoxic singlet oxygen ($^1$O$_2$). Competitively, \textbf{Ir$_1$} from the triplet state drops to the ground level radiatively, giving rise to phosphorescence (P).  \textcircled{2}  \textbf{Ir$_1$} molecules in the shell and in close proximity to gold core nanospheres leads to a resonant energy transfer process from Ir1 to the plasmonic nanostructure, with consequent conversion of the radiant energy into heat. The synergistic combination of photodynamic and photothermal effects induces fragmentation and reduction of the tumor vasculature network, preventing the delivery of oxygen and nutrients. The starved tumor regresses because of a massive devascularization leading to the complete tumor elimination.}
\label{fig:1}
\end{figure}

\section{Results}

 \textbf{Ir$_1$} and \textbf{Ir$_1$-AuSiO$_2$} were synthesized and characterized according to the reported procedures.\cite{ricciardi2017plasmon,ricciardi2014ionic} The results (Figures S1-S3, Supplementary Information) confirmed the successful preparations as reported in our previous study.\cite{ricciardi2017plasmon} 

Since PDT efficacy mostly depends on the PS ability to produce $^1$O$_2$, the amount generated by  \textbf{Ir$_1$}  and \textbf{Ir$_1$-AuSiO$_2$} was evaluated using 9,10-Anthracenediyl-bis (methylene)dimalonic acid (ABDA) as detection probe. The selected experimental conditions (PS concentration, fluence rate and distance between the excitation source and the sample solution), were the same used to carry out \textit{in vivo} studies, so as to mimic the $^1$O$_2$ generation occurring in the treated tumor tissue. Figure S4a shows the absorption spectra of ABDA in aqueous solution (control), in presence of  \textbf{Ir$_1$}  and \textbf{Ir$_1$-AuSiO$_2$} at different irradiation times. Although for the control sample no change in the ABDA optical density was observed, in presence of  \textbf{Ir$_1$}  or \textbf{Ir$_1$-AuSiO$_2$}, the ABDA absorption peaks decrease in intensity as the exposure time increases, highlighting a generation of $^1$O$_2$ in both cases but with different efficiencies. In fact, as clearly shown in Figure S4b, the plots of ABDA absorption at 378 nm as a function of exposure time, exhibit a linear trend with a slope for  \textbf{Ir$_1$}  sharper than that obtained for \textbf{Ir$_1$-AuSiO$_2$}. $^1$O$_2$ generation was quantitatively estimated calculating the total number of moles produced upon photoirradiation, obtaining values of 5.4$\cdot$10$^{-8}$ for \textbf{Ir$_1$} and 2.6$\cdot$10$^{-8}$ mol for \textbf{Ir$_1$-AuSiO$_2$} (Table S1). For comparison, the $^1$O$_2$ generation was measured in the same experimental conditions (PS concentration and radiation dose) for one of the most efficient phthalocyanine-based photosensitizer, the silicon phthalocyanine 4 (Pc4)\cite{zhao2009enhanced} (Table S1 and Figure S5). As reported in Table S1, the Pc4 exhibits a molar amount of $^1$O$_2$ generated of 4.3$\cdot$10$^{-8}$ moles, a value lower than that obtained for \textbf{Ir$_1$}.

\begin{figure}[H]
\includegraphics[width=1\columnwidth]{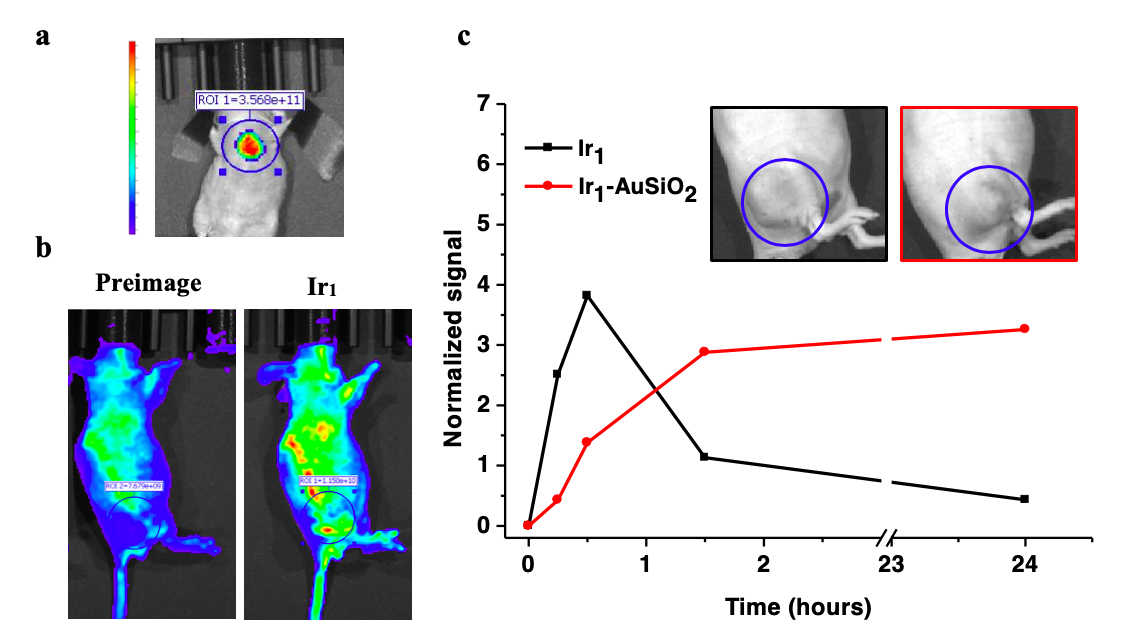}
\centering
\caption{\textbf{\textit{In vivo} imaging of mice treated with \textbf{Ir$_1$} or \textbf{Ir$_1$-AuSiO$_2$}.} \textbf{a}, Fluorescence image of  \textbf{Ir$_1$} injected subcutaneously into the mouse. The fluorescence signal intensity - measured as photon counts - is shown as color scale bar (radiant efficiency from 5.27$\cdot$10$^9$ (blue) to 4.74$\cdot$10$^{10}$ (red));  \textbf{b}, representative real-time \textit{in vivo} fluorescence images pre and 30 min post-injection of \textbf{Ir$_1$} into the GBM mass;  \textbf{c}, normalized signals acquired before time course and after injection of \textbf{Ir$_1$} (black) or \textbf{Ir$_1$-AuSiO$_2$} (red) into the GBM mass. Inset: bright field image of the flank tumor. The blue circle in all the images identifies the region-of-interest (ROI).
}
\label{fig:2}
\end{figure}

To demonstrate the efficiency of  \textbf{Ir$_1$} as luminescent probe for \textit{in vivo} imaging, we first tested the emission intensity - under 430 nm irradiation - immediately after a subcutaneous injection. As shown in Figure \ref{fig:2}a, the acquired fluorescence image exhibits - in the region-of-interest (ROI) - a huge emission. Then, we injected  \textbf{Ir$_1$} or \textbf{Ir$_1$-AuSiO$_2$} into the tumor of human GBM xenograft mouse model for real- time imaging at different time points post-injection, in order to follow their fate in terms of biodistribution, accumulation and permanence in the diseased region. Representative images pre and post-injection of  \textbf{Ir$_1$} are displayed in Figure \ref{fig:2}b; in particular, 30 min after the injection of  \textbf{Ir$_1$}, a bright fluorescence is highlighted in the whole body of the animal, widely spread, with a considerable intensity detected in the tumor region (ROI Figure \ref{fig:2}b). The fluorescence images acquired before the time course and after injection of both samples -  \textbf{Ir$_1$} and \textbf{Ir$_1$-AuSiO$_2$} - are reported in Figures S6 and S7, whereas the light signal intensity within the ROI as a function of time (0-24 h) is shown in Figure \ref{fig:2}c. Here, two distinct trends were observed. After injection of  \textbf{Ir$_1$}, the fluorescence intensity in the tumor region sharply increases, reaching a maximum after 30 min and then rapidly decreases within a few hours (Figures \ref{fig:2}c and S6). Conversely, the luminescent signal of \textbf{Ir$_1$-AuSiO$_2$} increases gradually up to 90 min, then remaining mostly constant even at 24 h post-injection (Figures \ref{fig:2}c and S7), suggesting a higher permanence of the luminescent nanostructures in the tumor tissue compared to the free photosensitizer.

\begin{figure}[H]
\includegraphics[width=0.6\columnwidth]{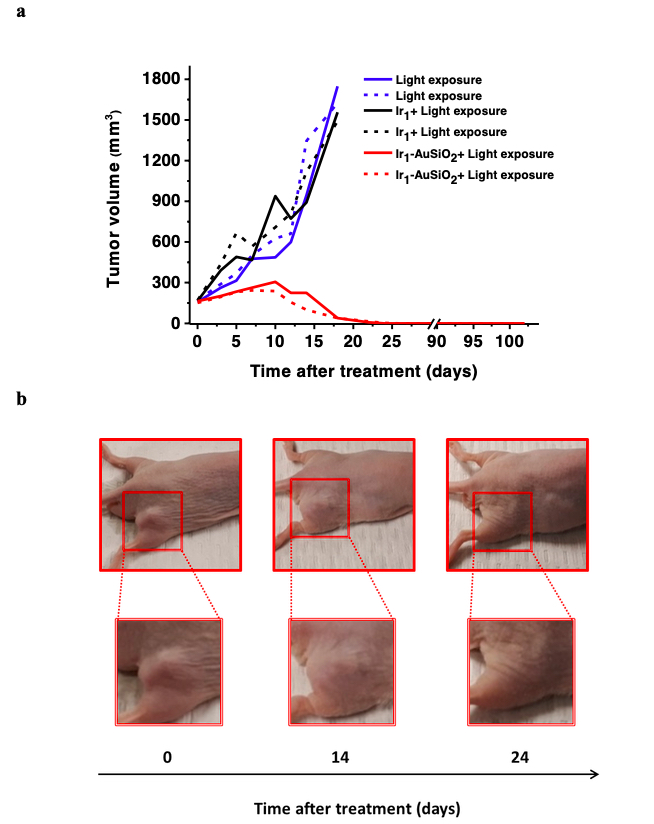}
\centering
\caption{\textbf{Follow-up of GBM xenografted mice after photo-treatment.}  \textbf{a}, Time-dependent tumor growth curves after light exposure (control group) or after intratumoral injection of \textbf{Ir$_1$} or \textbf{Ir$_1$-AuSiO$_2$} + light exposure (treatment groups), in GBM-bearing mice (n= 2 mice per group).  \textbf{b}, Photographs of a representative mouse from \textbf{Ir$_1$-AuSiO$_2$} + light exposure group showing the complete post treatment tumor regression.}
\label{fig:3}
\end{figure}

To evaluate the efficacy of  \textbf{Ir$_1$}  and \textbf{Ir$_1$-AuSiO$_2$} as phototherapeutic agents \textit{in vivo}, Gli36$\Delta$5 tumors were established into the right flank of nude mice.  \textbf{Ir$_1$}  or \textbf{Ir$_1$-AuSiO$_2$} were injected into the tumor site, then irradiated at 365 nm for 15 min (20 mW/cm$^2$), delivering a radiant exposure of 18 J/cm$^2$. As controls, one group received only the radiation dose and one group the intratumoral injection of  \textbf{Ir$_1$} or \textbf{Ir$_1$-AuSiO$_2$} without light exposure. Then, the phototherapeutic effect was assessed by monitoring the tumor volumes of the different groups over a period of 14 weeks. As clearly shown in Figures \ref{fig:3}a and S8 the control groups tumor sizes increased dramatically up to 1500 mm$^3$ in less than 18 days.
For group treated with \textbf{Ir$_1$} + light exposure, the tumor growth showed some fluctuations with spikes and slight inflections in the early days post treatment, and then a rapid increase, reaching the ethical limit (1500 mm$^3$) in 18 days (Figure \ref{fig:3}a). Noteworthy, for this treatment group, in the initial stage the growth of the tumor size is more significant than that observed for the control groups (Figures \ref{fig:3}a and S8). As regards the \textbf{Ir$_1$-AuSiO$_2$} treatment group, Figures \ref{fig:3}a and \ref{fig:3}b clearly show a volume increase of the tumor region in the phase immediately following the treatment, properly attributable to a swelling process, with a maximum value reached in 10 days. Beyond this stage, a gradual reduction of the tumor size was observed, leading to a surprisingly complete tumor regression in 24-26 days post treatment without highlighting relapses for over 100 days (Figure \ref{fig:3}a).

\begin{figure}[H]
\includegraphics[width=0.8\columnwidth]{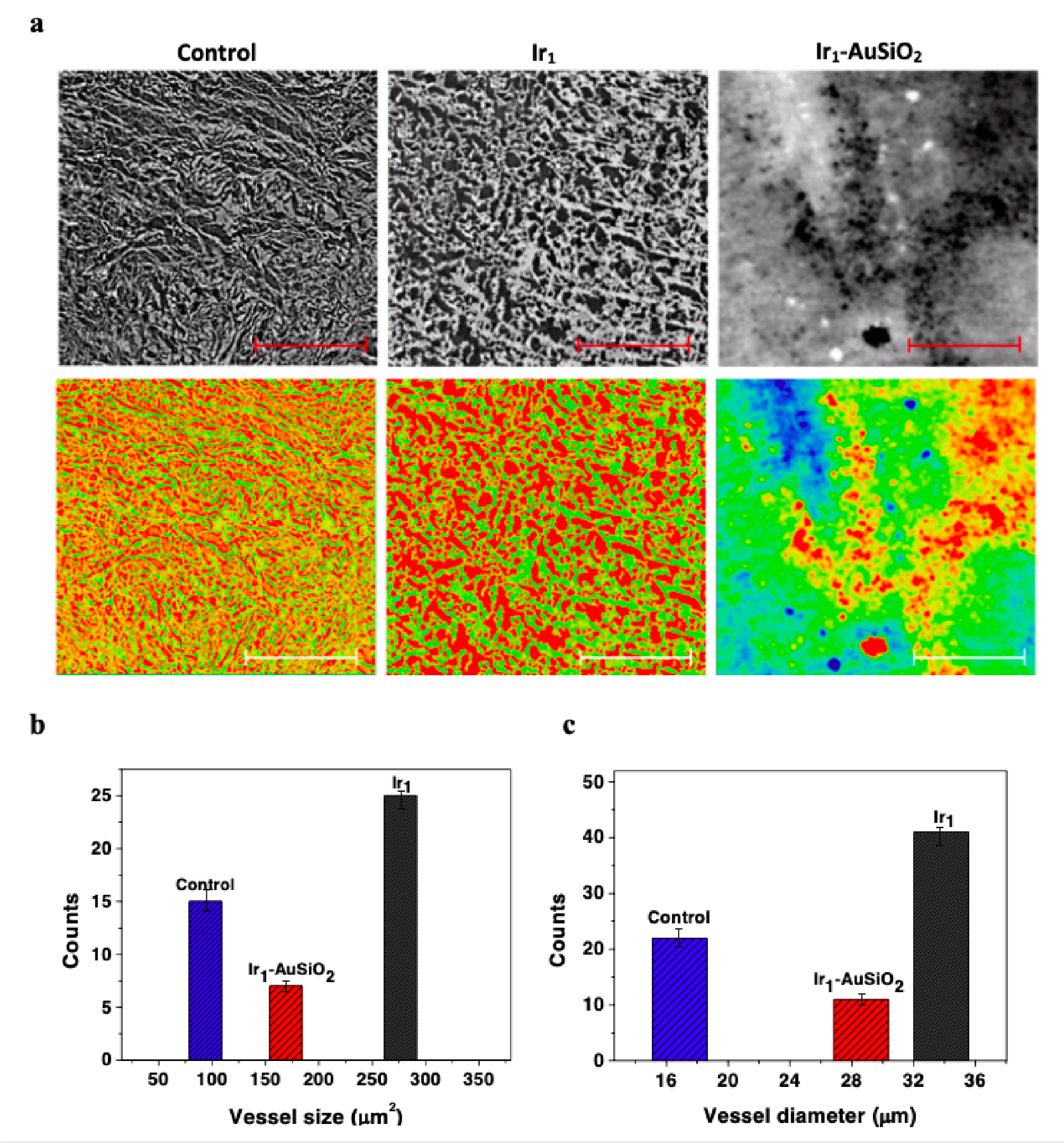}
\centering
\caption{\textbf{X-ray phase contrast tomography images and quantitative analysis of the vascular network of \textit{ex vivo} GBM samples.}  \textbf{a}, Grayscale and color tomography images of GBM samples treated - from left to right - only with the radiation dose (control), or after intratumoral injection of  \textbf{Ir$_1$} or \textbf{Ir$_1$-AuSiO$_2$} followed by light exposure. Red: blood vessels; green: GBM cells; blue: \textbf{Ir$_1$-AuSiO$_2$} clusters. Scale bar: 300 $\mu$m; box: 800$\times$800 pixels.  \textbf{b}, Average distribution of vessels at different size and c, diameter ranges in GBM samples treated only with the radiation dose (control), or after intratumoral injection of  \textbf{Ir$_1$} or \textbf{Ir$_1$-AuSiO$_2$} followed by light exposure. Quantitative analysis was performed on a working area of 400$\times$400 pixels.}
\label{fig:4}
\end{figure}

Figure \ref{fig:4}a shows tomographic images of  \textit{ex vivo} GBM samples - both control and treated - which were illuminated with the same light dose (365 nm for 15 min - 20 mW/cm$^2$) delivered during the \textit{in vivo} studies. The XPCT image of the control sample is reported as top left panel of Figure  \ref{fig:4}a, whereas the treated sample after intratumoral injection of  \textbf{Ir$_1$} or \textbf{Ir$_1$-AuSiO$_2$} are shown in the top-mid panel and top-right panel, respectively. Anatomical structures of GBM, such as blood vessels (black areas) and cancer cells (gray areas) are clearly recognized at micrometer-scale resolution in all the samples, whereas the sample treated with \textbf{Ir$_1$-AuSiO$_2$} shows clusters of nanoparticles (white areas) easily identified mostly in proximity of the microvessels. For better visualization, the same images are shown by using bright colors. The color gradient reflects the tissue density ranging from high (blue) to low (red) (Figure  \ref{fig:4}a, bottom). A quantitative analysis of vessel cross section ($\mu$m$^2$) and diameter ($\mu$m) is performed, and the average distribution is reported in Figures  \ref{fig:4}b and  \ref{fig:4}c, respectively. Both histograms are obtained by calculating a weighted average of the datasets displayed in Figures S9-S11. Regarding the control sample, an average lumen area and diameter of 94 $\mu$m$^2$ and 17 $\mu$m were respectively measured, in agreement with the values reported for GBM vessels.\cite{jain2007angiogenesis} By comparing the control sample with the tumor sample treated with only  \textbf{Ir$_1$}, a massive increase of the vascular network is clearly present, with a significantly larger vessel lumen area (276 $\mu$m$^2$) and diameter (34 $\mu$m) with respect to the control. On the contrary, the tumor treated with \textbf{Ir$_1$-AuSiO$_2$} shows a remarkable devascularization, in terms of a reduction in the number of microvessels. However, in this tumor sample we only observed a moderate increase in both the average vessel diameter (29 $\mu$m) and the average lumen area (169 $\mu$m$^2$).

\begin{figure}[H]
\includegraphics[width=0.7\columnwidth]{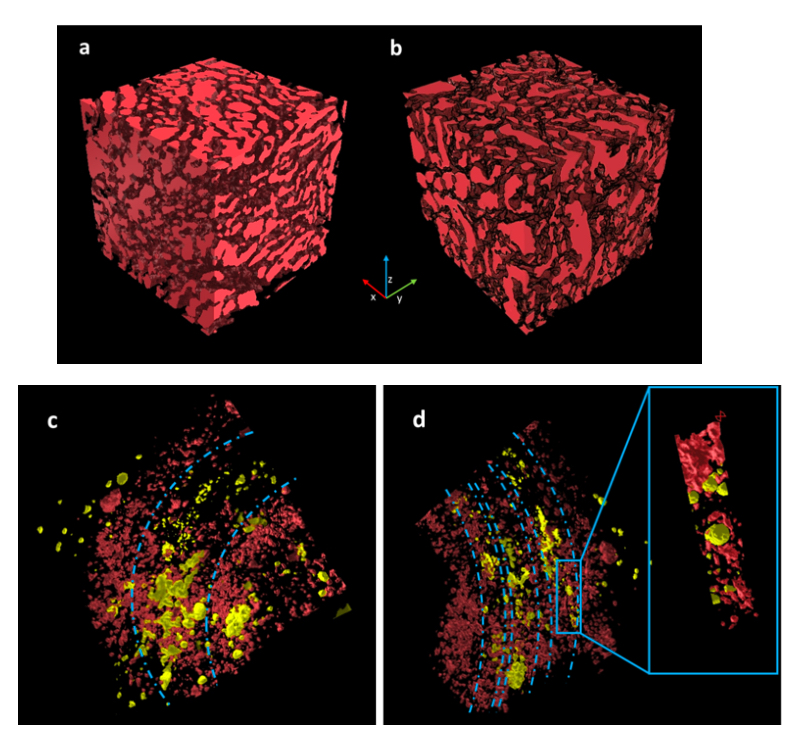}
\centering
\caption{\textbf{3D Images of the vasculature of \textit{ex vivo} GBM samples.}  3D volume (350$\times$350$\times$350 $\mu$m$^3$) of GBM samples treated \textbf{a}, only with the optical radiation dose (control) and \textbf{b}, after intratumoral injection of \textbf{Ir$_1$} followed by light exposure. The microvasculature is rendered in red. The tissues were computationally removed from the 3D rendering to highlight the vessels distribution. \textbf{c}, \textbf{d}, 3D volume (350$\times$350$\times$350 $\mu$m$^3$) of GBM sample after intratumoral injection of \textbf{Ir$_1$-AuSiO$_2$} followed by light exposure (top view). Vessels are rendered in red and the \textbf{Ir$_1$-AuSiO$_2$} clusters in yellow. Clusterized nanostructures appear distributed along some vascular channels highlighted by blue dashed lines. The inset highlights the details of nanoparticle clusters in the vascular structure and how their distribution correlates with the devascularized structure which appears to be reduced and discontinued.}
\label{fig:5}
\end{figure}

3D renderings of the segmented vascular network are reported in Figure  \ref{fig:5}. In particular, the control sample (Figure \ref{fig:5}a) shows an enormous hypervascularization of the tumor, where the microvascular structure appears as a glomeruloid tuft\cite{rojiani1996glomeruloid} with a high degree of branching. A similar network - with dense and proliferative vessels - is observed in the case of the GBM treated with \textbf{Ir$_1$} and exposed to lightwaves (Figure \ref{fig:5}b). On the contrary, as clearly shown in Figures \ref{fig:5}c, \ref{fig:5}d and Supplementary Movie 1, after \textbf{Ir$_1$-AuSiO$_2$}/light treatment the tumor tissue appears vastly devascularized, and the residual vascular structure seems to be collapsed, discontinuous and highly fragmented. Notably, the visible nanoparticles are not uniformly distributed within the examined tumor tissue; rather, they appear to be localized along preferential channels (Figures \ref{fig:5}c and \ref{fig:5}d).

\section{Discussion}

The ROS generation - in particular $^1$O$_2$ - is the key event underlying the tumor destruction process promoted by PDT agent and three distinct and interrelated mechanisms are responsible for the \textit{in vivo} anti-tumor effects of the phototherapeutic treatment.\cite{liu2019upconversion,castano2005mechanisms} In particular, the generated $^1$O$_2$ can (I) directly kill the tumor cells by apoptosis and/or necrosis, (II) damage the tumor-associated vasculature, resulting in tumor death via deprivation of oxygen and nutrients, (III) induce an acute inflammation with consequent activation of the immune response to recognize, track down and destroy tumor cells.\cite{castano2006photodynamic}

In our study, we performed a comparative analysis of the \textbf{Ir$_1$} and \textbf{Ir$_1$-AuSiO$_2$} $^1$O$_2$ generation efficiency for the experimental conditions chosen to carry out \textit{in vivo} studies. The high photosensitizing ability of the organometallic complex is observed by measuring the molar amount of $^1$O$_2$ generated after irradiation. Considering the solubility of molecular oxygen in water at room temperature (0.27 mmol/L),\cite{montalti2006handbook} we observe that after 15 minutes of light exposure in presence of \textbf{Ir$_1$}, most of the molecular oxygen present in solution was converted into $^1$O$_2$ (5.4$\cdot$10$^{-8}$ moles of 8.1$\cdot$10$^{-8}$, Table S1).

The \textbf{Ir$_1$-AuSiO$_2$} $^1$O$_2$ generation ability turns out to be lower than that of \textbf{Ir$_1$} at the same photosensitizer concentration. The spectral overlap between the emission band of the cyclometalated complex and the gold nanostructure plasmon resonance, leads to a competitive energy transfer process from the molecule to the metal core. \cite{ricciardi2017plasmon} Then, in the implemented nanosystem \textbf{Ir$_1$-AuSiO$_2$}, the radiant energy absorbed by \textbf{Ir$_1$} is partly transferred to molecular oxygen - generating $^1$O$_2$ - and partly to the metal core, with consequent conversion into heat.

\textit{In vivo} fluorescence imaging of GBM xenograft mouse model after intratumoral injection of \textbf{Ir$_1$} or \textbf{Ir$_1$-AuSiO$_2$}, allowed to track and localize over time the distribution of the PDT agent both as single molecules and embedded in the nanostructure. While the trafficking of \textbf{Ir$_1$} into the tumor site proves to be rapid, the luminescence intensity of \textbf{Ir$_1$-AuSiO$_2$} in the tumor reaches the maximum level with a slower rate, revealing a nanostructure permanence in the region of interest even 24 h post-injection. Since both \textbf{Ir$_1$} and \textbf{Ir$_1$-AuSiO$_2$} are water soluble, their size could be the key parameter dominating the different kinetics, thereby accumulating nanoparticles in the tumor region via EPR effect.\cite{golombek2018tumor,kalyane2019employment,shi2017cancer} Strikingly, the \textit{in vivo} treatment revealed opposite clinical outcomes: \textbf{Ir$_1$} treated tumors have shown a tumor growth similar to the control, whereas the cases treated with \textbf{Ir$_1$-AuSiO$_2$} have shown a slow regression which led to the tumor elimination without recurrence. 

To date, concerning the  \textit{in vivo} efficacy in GBM treatment, a decrease in the tumor growth and an increased survival of mice were observed by using several nanosystems, but no tumor eradication without relapse was observed.\cite{alphandery2020nano}

Upon analyzing the XPCT images after treatment with \textbf{Ir$_1$}, neovascularization accompanied by a marked vessel dilation is observed, whereas after \textbf{Ir$_1$-AuSiO$_2$}  treatment we observed a significant microvascular collapse with only a slight increase in the average diameter of the vessels. In order to elucidate the different effects caused by the bare \textbf{Ir$_1$} photosensitizer and by the \textbf{Ir$_1$-AuSiO$_2$}  nanosystem - in terms of mechanism of action and treatment outcome it has been important to focus on the peculiar features of the two samples. In both cases, the same light dose was applied few minutes after intratumoral injection. The short time frame between administration and light-mediated activation, excludes an efficient cell internalization of both phototherapeutic agents, \textbf{Ir$_1$} and even more \textbf{Ir$_1$-AuSiO$_2$}. This suggests that the localization may predominantly be in the interstitial space and/or close to the blood vessels, limiting a possible anti-tumor effect through the direct cellular damage of bulk tumor cells.

Regarding the process of generation of $^1$O$_2$,  \textbf{Ir$_1$} gives rise to an enormous amount of reactive oxygen species in a relatively short time (Figure S4 and Table S1). Paradoxically, a high $^1$O$_2$ generation rate achieved, for instance with high photosensitizer concentrations and/or high light fluence rate, causes faster oxygen consumption than that replaced by the bloodstream leading to severe levels of tumor hypoxia. The latter inducing the secretion of growth factors - has long been known as the major stimulator of tumor angiogenesis,\cite{castano2005mechanisms2} especially in GBM.\cite{jain2007angiogenesis,zagzag2000expression,shweiki1992vascular}

In our case, although \textbf{Ir$_1$} concentration and fluence rate are significantly low, the amount of $^1$O$_2$ generated in 15 min is high considering the total amount of molecular oxygen dissolved in solution. Conversely, the treatment carried out with  \textbf{Ir$_1$-AuSiO$_2$} exhibits a lower $^1$O$_2$ generation rate, avoiding plausibly oxygen-depleted conditions. Moreover, the sample treated with \textbf{Ir$_1$} highlights an average vessel dilation where approximately 60\% of vessels double the lumen area. The PDT-induced oxidative stress is known to activate an acute inflammatory response with consequent vessel dilation, which promotes the infiltration of the treated tissue by cells of the immune system with presentation of tumor-derived antigens.\cite{castano2006photodynamic}  As mentioned above, the anti-tumor immune development process is a potential tumor destruction mechanism mediated by PDT. However, we exclude an immune system response activated by these treatments since the \textit{in vivo} studies were carried out on immunodeficient nude mice. On the contrary, the higher blood supply, following the vessel dilation, could provide more nutrients 
and oxygen to the tumor, thus promoting its proliferation. These observations shed some light on our hypothesis regarding the failure of the \textit{in vivo} treatment only with \textbf{Ir$_1$}, as opposed to the results achieved with  \textbf{Ir$_1$-AuSiO$_2$}. In the latter case, in fact, a lower generation of reactive oxygen species determines a lower inflammatory response, as confirmed by the less pronounced vessel dilation.

\section{Conclusion}

To summarize, we report a comparative \textit{in vivo} study of GBM treatments by using two phototherapeutic agents: an Ir(III) complex (\textbf{Ir$_1$}) and the same compound embedded in core-shell plasmonic nanoparticles (\textbf{Ir$_1$-AuSiO$_2$}), spectrally resonant with the luminescent transition metal complex. GBM xenografted mice were exposed to a single light dose after treatment with the two therapeutics, inducing the generation of ROS only (\textbf{Ir$_1$}), or transducing the radiant energy into ROS and heat (\textbf{Ir$_1$-AuSiO$_2$}). The regression and elimination of the tumor in mice treated with \textbf{Ir$_1$-AuSiO$_2$} and exposed to VIS light, required a more detailed analysis of the damages to understand the mechanism underlying the tumor elimination. 3D high-resolution XPCT reveals a massive devascularization of the region where clusters of nanotherapeutic particles were concentrated, first inducing a vascular shut down followed by a significant tumor mass regression. On the other hand, XPCT study on mice treated with \textbf{Ir$_1$} shows that the vasculature remains structurally unmodified but a significant vessel dilation occurs as a consequence of local inflammatory processes. However, these results provide the evidence that nanotherapeutics based on plasmonic nanoparticles and weaponized with PDT agent have the potential to operate an effective killing physical mechanism in solid tumors by harnessing synergistic effects.

%%%%%%%%%%%%%%%%%%%%%%%%%%%%%%%%%%%%%%%%%%%%%%%%%%%%%%%%%%%%%%%%

\section{Associated Content}

Supplementary Information: Details of experimental materials and methods; characterization of the nanoparticles; comparative analysis of $^1$O$_2$generation detected by ABDA method; \textit{in vivo} imaging; X-ray phase contrast tomography images and quantitative analysis; Supplementary Movie.

%%%%%%%%%%%%%%%%%%%%%%%%%%%%%%%%%%%%%%%%%%%%%%%%%%%%%%%%%%%%%%%%

\begin{acknowledgement}

The authors thank Dr. Ramamurthy Gopalakrishnan and Prof. James P. Basilion (Case Western Reserve University, USA) for technical support in \textit{in vivo} studies. E.I.S. acknowledges the Romanian Academy, Program 4.

\end{acknowledgement}

%%%%%%%%%%%%%%%%%%%%%%%%%%%%%%%%%%%%%%%%%%%%%%%%%%%%%%%%%%%%%%%%
\section{Author contributions statement}

L.R., M.L. and G.S. conceived the research idea. L.R. carried out the nanoplatform synthesis/characterization and the singlet oxygen generation experiments. S.C. and L.R. carried out the treatments and contributed to acquire and analyze \textit{in vivo}  fluorescence imaging data in collaboration with Basilion's group. G.P. contributed to XPCT data processing. E.I.S., M.L. and L.R. carried out synthesis and characterization of the Ir(III) complex. A.S., F.P. N.P., M.F., R.B. and A.C. designed, carried out and analyzed the XPCT data. L.R. wrote the first draft of the manuscript. All authors discussed the results and provided constructive comments to the final manuscript. G.S. coordinated the research.

%%%%%%%%%%%%%%%%%%%%%%%%%%%%%%%%%%%%%%%%%%%%%%%%%%%%%%%%%%%%%%%%

\section{Competing interests}

The authors declare no competing financial interest.

%%%%%%%%%%%%%%%%%%%%%%%%%%%%%%%%%%%%%%%%%%%%%%%%%%%%%%%%%%%%%%%%

\bibliography{achemso-demo}

\providecommand{\latin}[1]{#1}
\makeatletter
\providecommand{\doi}
  {\begingroup\let\do\@makeother\dospecials
  \catcode`\{=1 \catcode`\}=2 \doi@aux}
\providecommand{\doi@aux}[1]{\endgroup\texttt{#1}}
\makeatother
\providecommand*\mcitethebibliography{\thebibliography}
\csname @ifundefined\endcsname{endmcitethebibliography}
  {\let\endmcitethebibliography\endthebibliography}{}
\begin{mcitethebibliography}{48}
\providecommand*\natexlab[1]{#1}
\providecommand*\mciteSetBstSublistMode[1]{}
\providecommand*\mciteSetBstMaxWidthForm[2]{}
\providecommand*\mciteBstWouldAddEndPuncttrue
  {\def\EndOfBibitem{\unskip.}}
\providecommand*\mciteBstWouldAddEndPunctfalse
  {\let\EndOfBibitem\relax}
\providecommand*\mciteSetBstMidEndSepPunct[3]{}
\providecommand*\mciteSetBstSublistLabelBeginEnd[3]{}
\providecommand*\EndOfBibitem{}
\mciteSetBstSublistMode{f}
\mciteSetBstMaxWidthForm{subitem}{(\alph{mcitesubitemcount})}
\mciteSetBstSublistLabelBeginEnd
  {\mcitemaxwidthsubitemform\space}
  {\relax}
  {\relax}

\bibitem[Louis \latin{et~al.}(2007)Louis, Ohgaki, Wiestler, Cavenee, Burger,
  Jouvet, Scheithauer, and Kleihues]{louis20072007}
Louis,~D.~N.; Ohgaki,~H.; Wiestler,~O.~D.; Cavenee,~W.~K.; Burger,~P.~C.;
  Jouvet,~A.; Scheithauer,~B.~W.; Kleihues,~P. The 2007 WHO classification of
  tumours of the central nervous system. \emph{Acta neuropathologica}
  \textbf{2007}, \emph{114}, 97--109\relax
\mciteBstWouldAddEndPuncttrue
\mciteSetBstMidEndSepPunct{\mcitedefaultmidpunct}
{\mcitedefaultendpunct}{\mcitedefaultseppunct}\relax
\EndOfBibitem
\bibitem[Schwartzbaum \latin{et~al.}(2006)Schwartzbaum, Fisher, Aldape, and
  Wrensch]{schwartzbaum2006epidemiology}
Schwartzbaum,~J.~A.; Fisher,~J.~L.; Aldape,~K.~D.; Wrensch,~M. Epidemiology and
  molecular pathology of glioma. \emph{Nature clinical practice Neurology}
  \textbf{2006}, \emph{2}, 494--503\relax
\mciteBstWouldAddEndPuncttrue
\mciteSetBstMidEndSepPunct{\mcitedefaultmidpunct}
{\mcitedefaultendpunct}{\mcitedefaultseppunct}\relax
\EndOfBibitem
\bibitem[Thakkar \latin{et~al.}(2014)Thakkar, Dolecek, Horbinski, Ostrom,
  Lightner, Barnholtz-Sloan, and Villano]{thakkar2014epidemiologic}
Thakkar,~J.~P.; Dolecek,~T.~A.; Horbinski,~C.; Ostrom,~Q.~T.; Lightner,~D.~D.;
  Barnholtz-Sloan,~J.~S.; Villano,~J.~L. Epidemiologic and molecular prognostic
  review of glioblastoma. \emph{Cancer Epidemiology and Prevention Biomarkers}
  \textbf{2014}, \emph{23}, 1985--1996\relax
\mciteBstWouldAddEndPuncttrue
\mciteSetBstMidEndSepPunct{\mcitedefaultmidpunct}
{\mcitedefaultendpunct}{\mcitedefaultseppunct}\relax
\EndOfBibitem
\bibitem[Hardee and Zagzag(2012)Hardee, and Zagzag]{hardee2012mechanisms}
Hardee,~M.~E.; Zagzag,~D. Mechanisms of glioma-associated neovascularization.
  \emph{The American journal of pathology} \textbf{2012}, \emph{181},
  1126--1141\relax
\mciteBstWouldAddEndPuncttrue
\mciteSetBstMidEndSepPunct{\mcitedefaultmidpunct}
{\mcitedefaultendpunct}{\mcitedefaultseppunct}\relax
\EndOfBibitem
\bibitem[Jain \latin{et~al.}(2007)Jain, Di~Tomaso, Duda, Loeffler, Sorensen,
  and Batchelor]{jain2007angiogenesis}
Jain,~R.~K.; Di~Tomaso,~E.; Duda,~D.~G.; Loeffler,~J.~S.; Sorensen,~A.~G.;
  Batchelor,~T.~T. Angiogenesis in brain tumours. \emph{Nature Reviews
  Neuroscience} \textbf{2007}, \emph{8}, 610--622\relax
\mciteBstWouldAddEndPuncttrue
\mciteSetBstMidEndSepPunct{\mcitedefaultmidpunct}
{\mcitedefaultendpunct}{\mcitedefaultseppunct}\relax
\EndOfBibitem
\bibitem[Stupp \latin{et~al.}(2009)Stupp, Hegi, Mason, Van Den~Bent, Taphoorn,
  Janzer, Ludwin, Allgeier, Fisher, Belanger, \latin{et~al.}
  others]{stupp2009effects}
Stupp,~R.; Hegi,~M.~E.; Mason,~W.~P.; Van Den~Bent,~M.~J.; Taphoorn,~M.~J.;
  Janzer,~R.~C.; Ludwin,~S.~K.; Allgeier,~A.; Fisher,~B.; Belanger,~K.,
  \latin{et~al.}  Effects of radiotherapy with concomitant and adjuvant
  temozolomide versus radiotherapy alone on survival in glioblastoma in a
  randomised phase \uppercase{III} study: 5-year analysis of the
  \uppercase{EORTC-NCIC} trial. \emph{The lancet oncology} \textbf{2009},
  \emph{10}, 459--466\relax
\mciteBstWouldAddEndPuncttrue
\mciteSetBstMidEndSepPunct{\mcitedefaultmidpunct}
{\mcitedefaultendpunct}{\mcitedefaultseppunct}\relax
\EndOfBibitem
\bibitem[Koshy \latin{et~al.}(2012)Koshy, Villano, Dolecek, Howard, Mahmood,
  Chmura, Weichselbaum, and McCarthy]{koshy2012improved}
Koshy,~M.; Villano,~J.~L.; Dolecek,~T.~A.; Howard,~A.; Mahmood,~U.;
  Chmura,~S.~J.; Weichselbaum,~R.~R.; McCarthy,~B.~J. Improved survival time
  trends for glioblastoma using the \uppercase{SEER} 17 population-based
  registries. \emph{Journal of neuro-oncology} \textbf{2012}, \emph{107},
  207--212\relax
\mciteBstWouldAddEndPuncttrue
\mciteSetBstMidEndSepPunct{\mcitedefaultmidpunct}
{\mcitedefaultendpunct}{\mcitedefaultseppunct}\relax
\EndOfBibitem
\bibitem[Tran and Rosenthal(2010)Tran, and Rosenthal]{tran2010survival}
Tran,~B.; Rosenthal,~M. Survival comparison between glioblastoma multiforme and
  other incurable cancers. \emph{Journal of Clinical Neuroscience}
  \textbf{2010}, \emph{17}, 417--421\relax
\mciteBstWouldAddEndPuncttrue
\mciteSetBstMidEndSepPunct{\mcitedefaultmidpunct}
{\mcitedefaultendpunct}{\mcitedefaultseppunct}\relax
\EndOfBibitem
\bibitem[Legler \latin{et~al.}(1999)Legler, Ries, Smith, Warren, Heineman,
  Kaplan, and Linet]{legler1999brain}
Legler,~J.~M.; Ries,~L. A.~G.; Smith,~M.~A.; Warren,~J.~L.; Heineman,~E.~F.;
  Kaplan,~R.~S.; Linet,~M.~S. Brain and other central nervous system cancers:
  recent trends in incidence and mortality. \emph{Journal of the National
  Cancer Institute} \textbf{1999}, \emph{91}, 1382--1390\relax
\mciteBstWouldAddEndPuncttrue
\mciteSetBstMidEndSepPunct{\mcitedefaultmidpunct}
{\mcitedefaultendpunct}{\mcitedefaultseppunct}\relax
\EndOfBibitem
\bibitem[De~Paula \latin{et~al.}(2017)De~Paula, Primo, and
  Tedesco]{de2017nanomedicine}
De~Paula,~L.~B.; Primo,~F.~L.; Tedesco,~A.~C. Nanomedicine associated with
  photodynamic therapy for glioblastoma treatment. \emph{Biophysical reviews}
  \textbf{2017}, \emph{9}, 761--773\relax
\mciteBstWouldAddEndPuncttrue
\mciteSetBstMidEndSepPunct{\mcitedefaultmidpunct}
{\mcitedefaultendpunct}{\mcitedefaultseppunct}\relax
\EndOfBibitem
\bibitem[Silbergeld and Chicoine(1997)Silbergeld, and
  Chicoine]{silbergeld1997isolation}
Silbergeld,~D.~L.; Chicoine,~M.~R. Isolation and characterization of human
  malignant glioma cells from histologically normal brain. \emph{Journal of
  neurosurgery} \textbf{1997}, \emph{86}, 525--531\relax
\mciteBstWouldAddEndPuncttrue
\mciteSetBstMidEndSepPunct{\mcitedefaultmidpunct}
{\mcitedefaultendpunct}{\mcitedefaultseppunct}\relax
\EndOfBibitem
\bibitem[Dupont \latin{et~al.}(2019)Dupont, Vermandel, Leroy, Quidet, Lecomte,
  Delhem, Mordon, and Reyns]{dupont2019intraoperative}
Dupont,~C.; Vermandel,~M.; Leroy,~H.-A.; Quidet,~M.; Lecomte,~F.; Delhem,~N.;
  Mordon,~S.; Reyns,~N. \uppercase{IN}traoperative photo\uppercase{DY}namic
  Therapy for \uppercase{G}li\uppercase{O}blastomas (\uppercase{INDYGO}): study
  protocol for a phase \uppercase{I} clinical trial. \emph{Neurosurgery}
  \textbf{2019}, \emph{84}, E414--E419\relax
\mciteBstWouldAddEndPuncttrue
\mciteSetBstMidEndSepPunct{\mcitedefaultmidpunct}
{\mcitedefaultendpunct}{\mcitedefaultseppunct}\relax
\EndOfBibitem
\bibitem[Dolmans \latin{et~al.}(2003)Dolmans, Fukumura, and
  Jain]{dolmans2003photodynamic}
Dolmans,~D.~E.; Fukumura,~D.; Jain,~R.~K. Photodynamic therapy for cancer.
  \emph{Nature reviews cancer} \textbf{2003}, \emph{3}, 380--387\relax
\mciteBstWouldAddEndPuncttrue
\mciteSetBstMidEndSepPunct{\mcitedefaultmidpunct}
{\mcitedefaultendpunct}{\mcitedefaultseppunct}\relax
\EndOfBibitem
\bibitem[Castano \latin{et~al.}(2004)Castano, Demidova, and
  Hamblin]{castano2004mechanisms}
Castano,~A.~P.; Demidova,~T.~N.; Hamblin,~M.~R. Mechanisms in photodynamic
  therapy: part one - photosensitizers, photochemistry and cellular
  localization. \emph{Photodiagnosis and photodynamic therapy} \textbf{2004},
  \emph{1}, 279--293\relax
\mciteBstWouldAddEndPuncttrue
\mciteSetBstMidEndSepPunct{\mcitedefaultmidpunct}
{\mcitedefaultendpunct}{\mcitedefaultseppunct}\relax
\EndOfBibitem
\bibitem[McKenzie \latin{et~al.}(2019)McKenzie, Bryant, and
  Weinstein]{mckenzie2019transition}
McKenzie,~L.~K.; Bryant,~H.~E.; Weinstein,~J.~A. Transition metal complexes as
  photosensitisers in one-and two-photon photodynamic therapy.
  \emph{Coordination Chemistry Reviews} \textbf{2019}, \emph{379}, 2--29\relax
\mciteBstWouldAddEndPuncttrue
\mciteSetBstMidEndSepPunct{\mcitedefaultmidpunct}
{\mcitedefaultendpunct}{\mcitedefaultseppunct}\relax
\EndOfBibitem
\bibitem[Zhao \latin{et~al.}(2013)Zhao, Wu, Sun, and Guo]{zhao2013triplet}
Zhao,~J.; Wu,~W.; Sun,~J.; Guo,~S. Triplet photosensitizers: from molecular
  design to applications. \emph{Chemical Society Reviews} \textbf{2013},
  \emph{42}, 5323--5351\relax
\mciteBstWouldAddEndPuncttrue
\mciteSetBstMidEndSepPunct{\mcitedefaultmidpunct}
{\mcitedefaultendpunct}{\mcitedefaultseppunct}\relax
\EndOfBibitem
\bibitem[Baggaley \latin{et~al.}(2012)Baggaley, Weinstein, and
  Williams]{baggaley2012lighting}
Baggaley,~E.; Weinstein,~J.~A.; Williams,~J.~G. Lighting the way to see inside
  the live cell with luminescent transition metal complexes. \emph{Coordination
  Chemistry Reviews} \textbf{2012}, \emph{256}, 1762--1785\relax
\mciteBstWouldAddEndPuncttrue
\mciteSetBstMidEndSepPunct{\mcitedefaultmidpunct}
{\mcitedefaultendpunct}{\mcitedefaultseppunct}\relax
\EndOfBibitem
\bibitem[Ko \latin{et~al.}(2019)Ko, Li, Leung, and Ma]{ko2019dual}
Ko,~C.-N.; Li,~G.; Leung,~C.-H.; Ma,~D.-L. Dual function luminescent transition
  metal complexes for cancer theranostics: The combination of diagnosis and
  therapy. \emph{Coordination Chemistry Reviews} \textbf{2019}, \emph{381},
  79--103\relax
\mciteBstWouldAddEndPuncttrue
\mciteSetBstMidEndSepPunct{\mcitedefaultmidpunct}
{\mcitedefaultendpunct}{\mcitedefaultseppunct}\relax
\EndOfBibitem
\bibitem[Monro \latin{et~al.}(2019)Monro, Col\'{o}n, Yin, Roque~III, Konda,
  Gujar, Thummel, Lilge, Cameron, and McFarland]{monro2018transition}
Monro,~S.; Col\'{o}n,~K.~L.; Yin,~H.; Roque~III,~J.; Konda,~P.; Gujar,~S.;
  Thummel,~R.~P.; Lilge,~L.; Cameron,~C.~G.; McFarland,~S.~A. Transition metal
  complexes and photodynamic therapy from a tumor-centered approach:
  Challenges, opportunities, and highlights from the development of
  \uppercase{TLD}1433. \emph{Chemical reviews} \textbf{2019}, \emph{119},
  797--828\relax
\mciteBstWouldAddEndPuncttrue
\mciteSetBstMidEndSepPunct{\mcitedefaultmidpunct}
{\mcitedefaultendpunct}{\mcitedefaultseppunct}\relax
\EndOfBibitem
\bibitem[Norouzi \latin{et~al.}(2019)Norouzi, Amerian, and
  Atyabi]{norouzi2019clinical}
Norouzi,~M.; Amerian,~M.; Atyabi,~F. Clinical applications of nanomedicine in
  cancer therapy. \emph{Drug discovery today} \textbf{2019}, \relax
\mciteBstWouldAddEndPunctfalse
\mciteSetBstMidEndSepPunct{\mcitedefaultmidpunct}
{}{\mcitedefaultseppunct}\relax
\EndOfBibitem
\bibitem[Wong \latin{et~al.}(2020)Wong, Sena-Torralba, {\'A}lvarez-Diduk,
  Muthoosamy, and Merko{\c{c}}i]{wong2020nanomaterials}
Wong,~X.~Y.; Sena-Torralba,~A.; {\'A}lvarez-Diduk,~R.; Muthoosamy,~K.;
  Merko{\c{c}}i,~A. Nanomaterials for nanotheranostics: tuning their properties
  according to disease needs. \emph{ACS nano} \textbf{2020}, \emph{14},
  2585--2627\relax
\mciteBstWouldAddEndPuncttrue
\mciteSetBstMidEndSepPunct{\mcitedefaultmidpunct}
{\mcitedefaultendpunct}{\mcitedefaultseppunct}\relax
\EndOfBibitem
\bibitem[Giner-Casares \latin{et~al.}(2016)Giner-Casares, Henriksen-Lacey,
  Coronado-Puchau, and Liz-Marz{\'a}n]{giner2016inorganic}
Giner-Casares,~J.~J.; Henriksen-Lacey,~M.; Coronado-Puchau,~M.;
  Liz-Marz{\'a}n,~L.~M. Inorganic nanoparticles for biomedicine: where
  materials scientists meet medical research. \emph{Materials Today}
  \textbf{2016}, \emph{19}, 19--28\relax
\mciteBstWouldAddEndPuncttrue
\mciteSetBstMidEndSepPunct{\mcitedefaultmidpunct}
{\mcitedefaultendpunct}{\mcitedefaultseppunct}\relax
\EndOfBibitem
\bibitem[Bobo \latin{et~al.}(2016)Bobo, Robinson, Islam, Thurecht, and
  Corrie]{bobo2016nanoparticle}
Bobo,~D.; Robinson,~K.~J.; Islam,~J.; Thurecht,~K.~J.; Corrie,~S.~R.
  Nanoparticle-based medicines: a review of
  \uppercase{F}\uppercase{D}\uppercase{A} - approved materials and clinical
  trials to date. \emph{Pharmaceutical research} \textbf{2016}, \emph{33},
  2373--2387\relax
\mciteBstWouldAddEndPuncttrue
\mciteSetBstMidEndSepPunct{\mcitedefaultmidpunct}
{\mcitedefaultendpunct}{\mcitedefaultseppunct}\relax
\EndOfBibitem
\bibitem[Moreira \latin{et~al.}(2018)Moreira, Rodrigues, Reis, Costa, and
  Correia]{moreira2018gold}
Moreira,~A.~F.; Rodrigues,~C.~F.; Reis,~C.~A.; Costa,~E.~C.; Correia,~I.~J.
  Gold-core silica shell nanoparticles application in imaging and therapy: A
  review. \emph{Microporous and Mesoporous Materials} \textbf{2018},
  \emph{270}, 168--179\relax
\mciteBstWouldAddEndPuncttrue
\mciteSetBstMidEndSepPunct{\mcitedefaultmidpunct}
{\mcitedefaultendpunct}{\mcitedefaultseppunct}\relax
\EndOfBibitem
\bibitem[Louis and Pluchery(2017)Louis, and Pluchery]{catherine2017gold}
Louis,~C.; Pluchery,~O. \emph{Gold nanoparticles for physics, chemistry and
  biology}; World Scientific, 2017\relax
\mciteBstWouldAddEndPuncttrue
\mciteSetBstMidEndSepPunct{\mcitedefaultmidpunct}
{\mcitedefaultendpunct}{\mcitedefaultseppunct}\relax
\EndOfBibitem
\bibitem[Hutter and Fendler(2004)Hutter, and Fendler]{hutter2004exploitation}
Hutter,~E.; Fendler,~J.~H. Exploitation of localized surface plasmon resonance.
  \emph{Advanced materials} \textbf{2004}, \emph{16}, 1685--1706\relax
\mciteBstWouldAddEndPuncttrue
\mciteSetBstMidEndSepPunct{\mcitedefaultmidpunct}
{\mcitedefaultendpunct}{\mcitedefaultseppunct}\relax
\EndOfBibitem
\bibitem[Huang \latin{et~al.}(2008)Huang, Jain, El-Sayed, and
  El-Sayed]{huang2008plasmonic}
Huang,~X.; Jain,~P.~K.; El-Sayed,~I.~H.; El-Sayed,~M.~A. Plasmonic photothermal
  therapy (\uppercase{PPTT}) using gold nanoparticles. \emph{Lasers in medical
  science} \textbf{2008}, \emph{23}, 217\relax
\mciteBstWouldAddEndPuncttrue
\mciteSetBstMidEndSepPunct{\mcitedefaultmidpunct}
{\mcitedefaultendpunct}{\mcitedefaultseppunct}\relax
\EndOfBibitem
\bibitem[Jaque \latin{et~al.}(2014)Jaque, Maestro, Del~Rosal, Haro-Gonzalez,
  Benayas, Plaza, Rodriguez, and Sole]{jaque2014nanoparticles}
Jaque,~D.; Maestro,~L.~M.; Del~Rosal,~B.; Haro-Gonzalez,~P.; Benayas,~A.;
  Plaza,~J.; Rodriguez,~E.~M.; Sole,~J.~G. Nanoparticles for photothermal
  therapies. \emph{Nanoscale} \textbf{2014}, \emph{6}, 9494--9530\relax
\mciteBstWouldAddEndPuncttrue
\mciteSetBstMidEndSepPunct{\mcitedefaultmidpunct}
{\mcitedefaultendpunct}{\mcitedefaultseppunct}\relax
\EndOfBibitem
\bibitem[Hu \latin{et~al.}(2020)Hu, Huang, Duan, Fu, and
  Liu]{hu2020reprogramming}
Hu,~Q.; Huang,~Z.; Duan,~Y.; Fu,~Z.; Liu,~B. Reprogramming Tumor
  Microenvironment with Photothermal Therapy. \emph{Bioconjugate Chemistry}
  \textbf{2020}, \emph{31}, 1268--1278\relax
\mciteBstWouldAddEndPuncttrue
\mciteSetBstMidEndSepPunct{\mcitedefaultmidpunct}
{\mcitedefaultendpunct}{\mcitedefaultseppunct}\relax
\EndOfBibitem
\bibitem[Vivero-Escoto \latin{et~al.}(2012)Vivero-Escoto, Huxford-Phillips, and
  Lin]{vivero2012silica}
Vivero-Escoto,~J.~L.; Huxford-Phillips,~R.~C.; Lin,~W. Silica-based nanoprobes
  for biomedical imaging and theranostic applications. \emph{Chemical Society
  Reviews} \textbf{2012}, \emph{41}, 2673--2685\relax
\mciteBstWouldAddEndPuncttrue
\mciteSetBstMidEndSepPunct{\mcitedefaultmidpunct}
{\mcitedefaultendpunct}{\mcitedefaultseppunct}\relax
\EndOfBibitem
\bibitem[Ricciardi \latin{et~al.}(2017)Ricciardi, Sancey, Palermo, Termine,
  De~Luca, Szerb, Aiello, Ghedini, Strangi, and La~Deda]{ricciardi2017plasmon}
Ricciardi,~L.; Sancey,~L.; Palermo,~G.; Termine,~R.; De~Luca,~A.; Szerb,~E.~I.;
  Aiello,~I.; Ghedini,~M.; Strangi,~G.; La~Deda,~M. Plasmon-mediated cancer
  phototherapy: the combined effect of thermal and photodynamic processes.
  \emph{Nanoscale} \textbf{2017}, \emph{9}, 19279--19289\relax
\mciteBstWouldAddEndPuncttrue
\mciteSetBstMidEndSepPunct{\mcitedefaultmidpunct}
{\mcitedefaultendpunct}{\mcitedefaultseppunct}\relax
\EndOfBibitem
\bibitem[Ricciardi \latin{et~al.}(2014)Ricciardi, Mastropietro, Ghedini,
  La~Deda, and Szerb]{ricciardi2014ionic}
Ricciardi,~L.; Mastropietro,~T.~F.; Ghedini,~M.; La~Deda,~M.; Szerb,~E.~I.
  Ionic-pair effect on the phosphorescence of ionic iridium (\uppercase{III})
  complexes. \emph{Journal of Organometallic Chemistry} \textbf{2014},
  \emph{772}, 307--313\relax
\mciteBstWouldAddEndPuncttrue
\mciteSetBstMidEndSepPunct{\mcitedefaultmidpunct}
{\mcitedefaultendpunct}{\mcitedefaultseppunct}\relax
\EndOfBibitem
\bibitem[Zhang \latin{et~al.}(2020)Zhang, Jian, Sun, Chen, Lv, Sun, Zhao, Zhao,
  and Hu]{zhang2020high}
Zhang,~W.; Jian,~J.; Sun,~C.; Chen,~J.; Lv,~W.; Sun,~M.; Zhao,~Y.; Zhao,~Q.;
  Hu,~C. High-resolution 3D imaging of microvascular architecture in human
  glioma tissues using X-ray phase-contrast computed tomography as a potential
  adjunct to histopathology. \emph{International Journal of Imaging Systems and
  Technology} \textbf{2020}, \emph{30}, 464--472\relax
\mciteBstWouldAddEndPuncttrue
\mciteSetBstMidEndSepPunct{\mcitedefaultmidpunct}
{\mcitedefaultendpunct}{\mcitedefaultseppunct}\relax
\EndOfBibitem
\bibitem[Momose \latin{et~al.}(1996)Momose, Takeda, Itai, and
  Hirano]{momose1996phase}
Momose,~A.; Takeda,~T.; Itai,~Y.; Hirano,~K. Phase--contrast X--ray computed
  tomography for observing biological soft tissues. \emph{Nature medicine}
  \textbf{1996}, \emph{2}, 473--475\relax
\mciteBstWouldAddEndPuncttrue
\mciteSetBstMidEndSepPunct{\mcitedefaultmidpunct}
{\mcitedefaultendpunct}{\mcitedefaultseppunct}\relax
\EndOfBibitem
\bibitem[Zhao \latin{et~al.}(2009)Zhao, Yin, Bilski, Chignell, Roberts, and
  He]{zhao2009enhanced}
Zhao,~B.; Yin,~J.-J.; Bilski,~P.~J.; Chignell,~C.~F.; Roberts,~J.~E.; He,~Y.-Y.
  Enhanced photodynamic efficacy towards melanoma cells by encapsulation of
  \uppercase{P}c4 in silica nanoparticles. \emph{Toxicology and applied
  pharmacology} \textbf{2009}, \emph{241}, 163--172\relax
\mciteBstWouldAddEndPuncttrue
\mciteSetBstMidEndSepPunct{\mcitedefaultmidpunct}
{\mcitedefaultendpunct}{\mcitedefaultseppunct}\relax
\EndOfBibitem
\bibitem[Rojiani and Dorovini-Zis(1996)Rojiani, and
  Dorovini-Zis]{rojiani1996glomeruloid}
Rojiani,~A.~M.; Dorovini-Zis,~K. Glomeruloid vascular structures in
  glioblastoma multiforme: an immunohistochemical and ultrastructural study.
  \emph{Journal of neurosurgery} \textbf{1996}, \emph{85}, 1078--1084\relax
\mciteBstWouldAddEndPuncttrue
\mciteSetBstMidEndSepPunct{\mcitedefaultmidpunct}
{\mcitedefaultendpunct}{\mcitedefaultseppunct}\relax
\EndOfBibitem
\bibitem[Liu \latin{et~al.}(2019)Liu, Meng, and Bu]{liu2019upconversion}
Liu,~Y.; Meng,~X.; Bu,~W. Upconversion-based photodynamic cancer therapy.
  \emph{Coordination Chemistry Reviews} \textbf{2019}, \emph{379}, 82--98\relax
\mciteBstWouldAddEndPuncttrue
\mciteSetBstMidEndSepPunct{\mcitedefaultmidpunct}
{\mcitedefaultendpunct}{\mcitedefaultseppunct}\relax
\EndOfBibitem
\bibitem[Castano \latin{et~al.}(2005)Castano, Demidova, and
  Hamblin]{castano2005mechanisms}
Castano,~A.~P.; Demidova,~T.~N.; Hamblin,~M.~R. Mechanisms in photodynamic
  therapy: part three - photosensitizer pharmacokinetics, biodistribution,
  tumor localization and modes of tumor destruction. \emph{Photodiagnosis and
  photodynamic therapy} \textbf{2005}, \emph{2}, 91--106\relax
\mciteBstWouldAddEndPuncttrue
\mciteSetBstMidEndSepPunct{\mcitedefaultmidpunct}
{\mcitedefaultendpunct}{\mcitedefaultseppunct}\relax
\EndOfBibitem
\bibitem[Castano \latin{et~al.}(2006)Castano, Mroz, and
  Hamblin]{castano2006photodynamic}
Castano,~A.~P.; Mroz,~P.; Hamblin,~M.~R. Photodynamic therapy and anti-tumour
  immunity. \emph{Nature Reviews Cancer} \textbf{2006}, \emph{6},
  535--545\relax
\mciteBstWouldAddEndPuncttrue
\mciteSetBstMidEndSepPunct{\mcitedefaultmidpunct}
{\mcitedefaultendpunct}{\mcitedefaultseppunct}\relax
\EndOfBibitem
\bibitem[Montalti \latin{et~al.}(2006)Montalti, Credi, Prodi, and
  Gandolfi]{montalti2006handbook}
Montalti,~M.; Credi,~A.; Prodi,~L.; Gandolfi,~M.~T. \emph{Handbook of
  photochemistry}; CRC press, 2006\relax
\mciteBstWouldAddEndPuncttrue
\mciteSetBstMidEndSepPunct{\mcitedefaultmidpunct}
{\mcitedefaultendpunct}{\mcitedefaultseppunct}\relax
\EndOfBibitem
\bibitem[Golombek \latin{et~al.}(2018)Golombek, May, Theek, Appold, Drude,
  Kiessling, and Lammers]{golombek2018tumor}
Golombek,~S.~K.; May,~J.-N.; Theek,~B.; Appold,~L.; Drude,~N.; Kiessling,~F.;
  Lammers,~T. Tumor targeting via \uppercase{EPR}: Strategies to enhance
  patient responses. \emph{Advanced drug delivery reviews} \textbf{2018},
  \emph{130}, 17--38\relax
\mciteBstWouldAddEndPuncttrue
\mciteSetBstMidEndSepPunct{\mcitedefaultmidpunct}
{\mcitedefaultendpunct}{\mcitedefaultseppunct}\relax
\EndOfBibitem
\bibitem[Kalyane \latin{et~al.}(2019)Kalyane, Raval, Maheshwari, Tambe, Kalia,
  and Tekade]{kalyane2019employment}
Kalyane,~D.; Raval,~N.; Maheshwari,~R.; Tambe,~V.; Kalia,~K.; Tekade,~R.~K.
  Employment of enhanced permeability and retention effect
  (\uppercase{E}\uppercase{P}\uppercase{R}): Nanoparticle-based precision tools
  for targeting of therapeutic and diagnostic agent in cancer. \emph{Materials
  Science and Engineering: C} \textbf{2019}, \emph{98}, 1252--1276\relax
\mciteBstWouldAddEndPuncttrue
\mciteSetBstMidEndSepPunct{\mcitedefaultmidpunct}
{\mcitedefaultendpunct}{\mcitedefaultseppunct}\relax
\EndOfBibitem
\bibitem[Shi \latin{et~al.}(2017)Shi, Kantoff, Wooster, and
  Farokhzad]{shi2017cancer}
Shi,~J.; Kantoff,~P.~W.; Wooster,~R.; Farokhzad,~O.~C. Cancer nanomedicine:
  progress, challenges and opportunities. \emph{Nature Reviews Cancer}
  \textbf{2017}, \emph{17}, 20\relax
\mciteBstWouldAddEndPuncttrue
\mciteSetBstMidEndSepPunct{\mcitedefaultmidpunct}
{\mcitedefaultendpunct}{\mcitedefaultseppunct}\relax
\EndOfBibitem
\bibitem[Alphand{\'e}ry(2020)]{alphandery2020nano}
Alphand{\'e}ry,~E. Nano-Therapies for Glioblastoma Treatment. \emph{Cancers}
  \textbf{2020}, \emph{12}, 242\relax
\mciteBstWouldAddEndPuncttrue
\mciteSetBstMidEndSepPunct{\mcitedefaultmidpunct}
{\mcitedefaultendpunct}{\mcitedefaultseppunct}\relax
\EndOfBibitem
\bibitem[Castano \latin{et~al.}(2005)Castano, Demidova, and
  Hamblin]{castano2005mechanisms2}
Castano,~A.~P.; Demidova,~T.~N.; Hamblin,~M.~R. Mechanisms in photodynamic
  therapy: part two - cellular signaling, cell metabolism and modes of cell
  death. \emph{Photodiagnosis and photodynamic therapy} \textbf{2005},
  \emph{2}, 1--23\relax
\mciteBstWouldAddEndPuncttrue
\mciteSetBstMidEndSepPunct{\mcitedefaultmidpunct}
{\mcitedefaultendpunct}{\mcitedefaultseppunct}\relax
\EndOfBibitem
\bibitem[Zagzag \latin{et~al.}(2000)Zagzag, Zhong, Scalzitti, Laughner, Simons,
  and Semenza]{zagzag2000expression}
Zagzag,~D.; Zhong,~H.; Scalzitti,~J.~M.; Laughner,~E.; Simons,~J.~W.;
  Semenza,~G.~L. Expression of hypoxia-inducible factor 1$\alpha$ in brain
  tumors: association with angiogenesis, invasion, and progression.
  \emph{Cancer: Interdisciplinary International Journal of the American Cancer
  Society} \textbf{2000}, \emph{88}, 2606--2618\relax
\mciteBstWouldAddEndPuncttrue
\mciteSetBstMidEndSepPunct{\mcitedefaultmidpunct}
{\mcitedefaultendpunct}{\mcitedefaultseppunct}\relax
\EndOfBibitem
\bibitem[Shweiki \latin{et~al.}(1992)Shweiki, Itin, Soffer, and
  Keshet]{shweiki1992vascular}
Shweiki,~D.; Itin,~A.; Soffer,~D.; Keshet,~E. Vascular endothelial growth
  factor induced by hypoxia may mediate hypoxia-initiated angiogenesis.
  \emph{Nature} \textbf{1992}, \emph{359}, 843--845\relax
\mciteBstWouldAddEndPuncttrue
\mciteSetBstMidEndSepPunct{\mcitedefaultmidpunct}
{\mcitedefaultendpunct}{\mcitedefaultseppunct}\relax
\EndOfBibitem
\end{mcitethebibliography}

\end{document}